\documentclass[12pt]{spieman}
\usepackage{setspace}

\usepackage[english]{babel}
\usepackage[utf8]{inputenc}

\usepackage{amsmath,amsfonts,amssymb}
\usepackage{bm}
\usepackage{siunitx}
\usepackage{newtxmath}

\usepackage{graphicx}
\usepackage[section]{placeins}
\usepackage{caption}
\usepackage{subcaption}
\usepackage{makecell}

\usepackage{etoolbox}
\usepackage[title]{appendix}
\AtBeginEnvironment{appendices}{\crefalias{section}{appendix}}

\usepackage{tocloft}

\cftpagenumbersoff{figure}
\cftpagenumbersoff{table} 

\usepackage{hyperref}
\usepackage{cleveref}

\newcommand{\ax}{\vartheta_x}
\newcommand{\ay}{\vartheta_y}
\newcommand{\axm}{\vartheta_{x,m}}
\newcommand{\aym}{\vartheta_{y,m}}

\newcommand{\px}{p_x}
\newcommand{\py}{p_y}
\newcommand{\pxm}{p_{x,m}}
\newcommand{\pym}{p_{y,m}}

\newcommand{\cur}{c}
\newcommand{\curm}{\cur_m}

\newcommand{\ds}{D4\sigma}
\newcommand{\dsm}{\ds_m}

\title{Analysis and Mitigation of Crosstalk in a Multi-Parameter Laser Beam Diagnostic System}
\author[1,*]{Benjamin Nagler}
\affil[1]{Hensoldt Optronics GmbH\\Carl-Zeiss-Straße 22\\73447 Oberkochen\\Germany}

\begin{document}
	\maketitle
	
	\begin{abstract}
		Accurate characterization of laser beams is crucial for applications ranging from laser-based material processing to gravitational wave detection, where even minor measurement deviations can significantly impact system performance. This work investigates crosstalk effects in a multi-parameter laser beam diagnostic system designed to simultaneously measure six beam parameters: pointing angles, centroid positions, wavefront curvature, and beam diameter. Using sequential ray-optical simulations in Zemax, a comprehensive analysis of parameter interdependencies is performed through a systematic parameter sweep across 15625 unique combinations within specified measurement ranges. The results reveal significant crosstalk effects between several parameters, which are traced back to specific optical components in the system. To address these issues and improve accuracy, a mathematical framework for crosstalk mitigation based on Taylor series expansions is developed. The mitigation method reduces measurement deviations significantly by up to one order of magnitude. This work provides insights into the origins and mitigation of crosstalk effects in optical measurement systems, particularly relevant for high-accuracy applications.
	\end{abstract}


    {\noindent \footnotesize\textbf{*}Benjamin Nagler, \linkable{benjamin.nagler@hensoldt.net}}

    \begin{spacing}{1}
    
	\section{Introduction}
        Modern laser-based technologies increasingly demand unprecedented levels of beam parameter measurement and control. In gravitational wave detectors like LIGO~\cite{instruments6010015}, beam pointing stability must be maintained at the \si{\nano\radian} level to achieve the required measurement sensitivity~\cite{ligo2015}. Similarly, laser material processing requires precise beam control for consistent results~\cite{understandinglaser}, while laser-based fusion facilities depend on accurate beam characterization for optimal energy delivery~\cite{Boege01111998}. In state-of-the-art semiconductor manufacturing, accurate laser control is crucial for generating the extreme ultraviolet light used in photolithography, where beam stability affects the quality of produced chips~\cite{doi:10.1515/aot-2017-0029}. Laser beam diagnostic systems typically measure multiple parameters simultaneously~\cite{10.1117/12.810515,10.1117/12.639286,iso11146-2}, including both intensity-based characteristics (e.g., beam centroid position and beam diameter) and properties of the wavefront (e.g., pointing angle, curvature, astigmatism).
		
		Crosstalk in a measurement device emerges when the variation of one measured parameter distorts the measurement of another parameter, even though these parameters are inherently uncorrelated outside the measurement process.~\cite{mazda1993telecommunications,wiki:crosstalk}. It presents a fundamental challenge by creating deceptive dependencies that compromise accuracy. This phenomenon manifests itself across multiple domains, with, for example, electronic systems experiencing issues like coupling effects between adjacent circuit elements~\cite{crosstalkdigital} and heat conduction between components~\cite{crosstalkthermal}. For the emerging field of quantum computing, crosstalk remains a major challenge affecting accuracy and performance~\cite{Sarovar2020detectingcrosstalk}.  In optical systems, crosstalk may be caused by stray light~\cite{crosstalkstray}, insufficient isolation between measurement channels~\cite{crosstalkapdarray}, non-linear optical effects~\cite{monroy2013crosstalk}, and optical aberrations~\cite{Hecht2018} - with the latter being particularly significant in optical imaging systems. Understanding and mitigating such crosstalk is essential for achieving accurate measurements in many measurement systems.
        
		In laser beam diagnostic systems, the impact of crosstalk depends on how the system is implemented and integrated. When the diagnostic system, such as a beam profiler or a wavefront sensor, is placed directly into the beam path, crosstalk is often less crucial because these systems do not rely on additional optical elements to sample or image the beam. In inline diagnostic systems, however, where laser beam parameters must be measured at a specific position along the beam path without obstructing it, typically a fraction of the beam power is split off for analysis. This usually requires imaging of the beam at the sampled position onto the sensor~\cite{10.1117/12.810515}. The imaging system itself can introduce crosstalk through optical aberrations, tolerances of optical and opto-mechanical components, apertures partially trimming the beam, or imperfect alignment, which can couple different measured beam parameters.
		
		As an illustrative  example, consider a lens with focal length $f$ used for angle measurement by focusing the beam and measuring the lateral spot centroid position with a position sensor~\cite{7234874}. When the location of the position sensor coincides with the focal plane, the measured spot position $p_m=f\vartheta$ (in paraxial approximation) depends solely on the incident angle $\vartheta$. If, however, the beam acquires a quadratic phase term (Zernike defocus~\cite{Goodwin2006-vy}) with curvature $\cur$, possibly due to imperfectly aligned imaging optics, the measurement becomes position dependent. This occurs because the curvature induces a focal plane shift of $\Delta f = \cur f^2$, resulting in a measured position of $p_m = (f + \Delta f)\vartheta - p_0\Delta f/f$, where $p_0$ is the initial beam position. This coupling causes crosstalk between the beam position $p_0$ and the measured beam pointing angle $\vartheta_m=p_m/f$. The magnitude and direction of this linear crosstalk can be quantified by the first derivative $\mathrm{d}\vartheta_m/\mathrm{d} p_0 =-\cur$, which, as expected, vanishes for zero curvature.
		
        This article presents a comprehensive crosstalk analysis of a multi-parameter laser beam diagnostic system using ray-optical simulations. The investigation reveals how variations in one beam parameter influence the measurement of others, examining all possible combinations of beam pointing angle, centroid position, wavefront curvature, and diameter through systematic parameter sweeps. A detailed analysis identifies the physical origins of the observed crosstalk effects, tracing them back to specific optical components in the system. Based on these findings, a mathematical framework using Taylor series expansions is developed that describes the dependencies between measured and set parameters through a system of equations. Solving these equations for the set parameters enables mitigation of crosstalk effects, with significant improvements in measurement accuracy demonstrated across the full parameter space.

    \section{Materials and Methods}
    	\subsection{Description of the beam diagnostic system}
             \begin{figure}[t]
                \centering
                \includegraphics[width=0.9\textwidth]{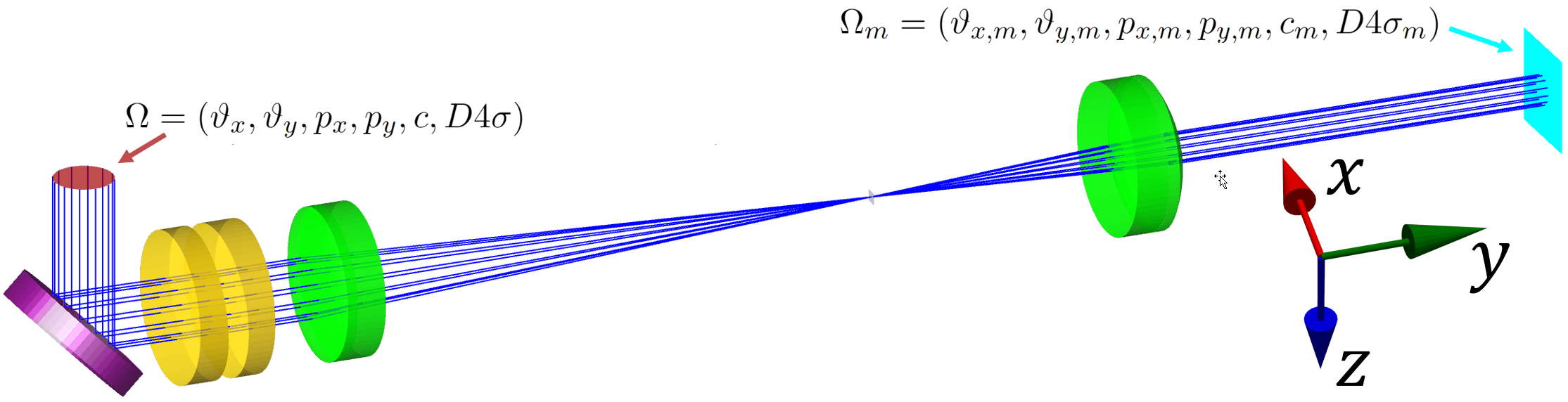}
                \caption[Optical setup of the laser beam diagnostic system and coordinate system definition]{Optical setup of the laser beam diagnostic system and coordinate system definition. The measurement plane (red) is the plane where the laser beam parameters (set values $\Omega$) are to be determined. The origin of the coordinate system is located in the center of the measurement plane, which coincides with the $x$-$y$-plane. A wedged beam splitter (purple) picks up a small portion of the beam power and reflects it towards the imaging system. A pair of rotating wedges (yellow) eliminate residual pointing angles, while the Kepler telescope (green) images the measurement plane onto the sensor. Stray light from ghost reflections off the wedged beam splitter's backside is blocked by a field stop (gray) in the focal plane of the first Kepler lens. The measured values of the beam parameters ($\Omega_m$) are obtained in the wavefront sensor plane (bright blue). Lenses and wedge plates have a diameter of \SI{25.4}{\milli\meter}, the displayed ray bundle has a diameter of \SI{10}{\milli\meter}. A detailed description of the optical setup is provided in  \Cref{sec:app:opticalsystem}.}
                \label{fig:system_layout}
             \end{figure}     
        
            \Cref{fig:system_layout} shows the custom-built inline beam diagnostic system. The laser beam parameters are measured at a designated measurement plane, which is imaged onto a wavefront sensor by a Kepler telescope with magnification \mbox{$M=0.5$}. The telescope employs two commercially available achromatic lens doublets (Thorlabs AC254-100-C and AC254-050-C with focal lengths of \SI{100}{\milli\meter} and \SI{50}{\milli\meter}, respectively). 
    
            A wedged plate beam splitter (wedge angle \SI{20}{\milli\radian}) redirects a small fraction ($<\SI{1}{\percent}$) of the incident beam power into the diagnostic beam path while transmitting the majority power for subsequent applications. The wedged front surface of the beam splitter imparts an additional angle to spurious ghost reflections from its backside, enabling their removal at the field stop (rectangular aperture with size $\SI{2}{\milli\meter} \times \SI{2}{\milli\meter}$) located in the focal point of the telescope.
            
            To compensate for the beam deflection caused by reflection off the wedged front surface of the beam splitter, the system incorporates a pair of rotating wedges with \SI{50}{\milli\radian} wedge angles. These wedges serve a dual purpose: they both neutralize the deflection angle and enable correction of incoming beam offset angles within the specified range. Importantly, the wedges also cause a lateral displacement of transmitted rays.
            
            The system might utilize a multi-wave lateral shearing interferometer~\cite{Chanteloup2005me} or a Shack-Hartmann sensor~\cite{Platt2001HistoryAP} for wavefront sensing, both of which measure the beam's wavefront and its transversal intensity distribution~\cite{Schafer:02,Mikhaylov2020-ci}. Since this investigation focuses on crosstalk caused by the optical imaging system, the details of the wavefront sensor are not considered in the simulation.
    
            The system is optimized to operate in a broad wavelength range between \SI{1064}{\nano\meter} and \SI{1550}{\nano\meter}. This optimization encompasses multiple system parameters, including the rotating wedge angles, the telescope lens separation, and the distance between the second telescope lens and the wavefront sensor. Due to the inherent constraints of broadband optimization, perfect alignment cannot be achieved across all wavelengths within the specified range. These wavelength-dependent deviations from ideal alignment constitute a significant source of measurement crosstalk, as demonstrated in later sections. In practical implementations, manufacturing tolerances of optical components and limitations in the opto-mechanical assembly can introduce additional crosstalk effects. A detailed description of the optical setup is provided in \Cref{sec:app:opticalsystem}.
            
            The system is designed to measure six fundamental parameters of a Gaussian laser beam~\cite{Mandel_Wolf_1995} in the measurement plane: the pointing angles $\ax$ and $\ay$ along two orthogonal axes $x$ and $y$, the lateral intensity profile centroid positions $\px$ and $\py$ along $x$ and $y$, the wavefront curvature $\cur$, and the laser beam diameter $\ds$~\cite{iso11146-1}. In the following, the ensembles of set and measured parameters are denoted as $\Omega=(\ax,\ay,\px,\py,\cur,\ds)$ and $\Omega_m=(\axm,\aym,\pxm,\pym,\curm,\dsm)$, respectively. The system has been specifically designed to accommodate the parameter measurement ranges listed in \Cref{tab:measurement_ranges}, which form the basis for the subsequent analysis. The nominal values listed in the table are the ones used for the optimization.
            
            \begin{table}[htbp]
                \centering
                \caption{Parameter measurement ranges}
                \label{tab:measurement_ranges}
                \begin{tabular}{lcccc}
                    \hline
                    parameter & \makecell{symbol\\set value} & \makecell{symbol\\measured value} & \makecell{measurement\\range} & \makecell{nominal\\value} \\
                    \hline
                    pointing angle $x$ & $\ax$ & $\axm$ & \SI{\pm 5}{\milli\radian} & \SI{0}{\milli\radian} \\
                    pointing angle $y$ & $\ay$ & $\aym$ & \SI{\pm 5}{\milli\radian} & \SI{0}{\milli\radian} \\
                    centroid position $x$ & $\px$ & $\pxm$ & \SI{\pm 1}{\milli\meter} & \SI{0}{\milli\meter} \\
                    centroid position $y$ &  $\py$ & $\pym$ & \SI{\pm 1}{\milli\meter} & \SI{0}{\milli\meter} \\
                    wavefront curvature & $\cur$ & $\curm$ & \SI{\pm 0.03}{\per\meter} & \SI{0}{\per\meter} \\
                    beam diameter & $\ds$ & $\dsm$ & \SIrange{3.27}{8.17}{\milli\meter} & \SI{5.72}{\milli\meter} \\
                    \hline
                \end{tabular}
            \end{table}
    			
    	\subsection{Optical simulation of the beam diagnostic system}   
\begin{figure}[t]
    			\centering
    			\includegraphics[width=1\textwidth]{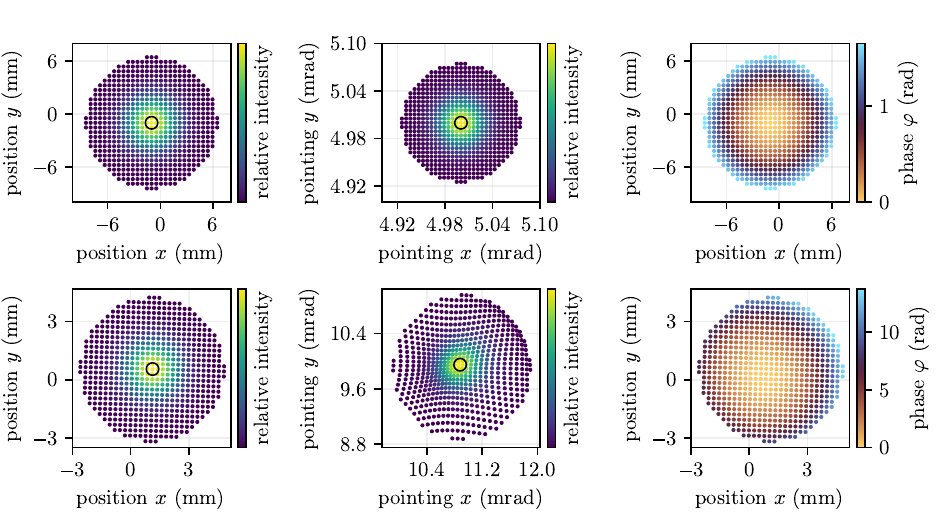}
    			\caption[Exemplary intensity and phase distributions]{Exemplary intensity (in position space and angle space) and phase distributions in the measurement plane (upper plots) and at the wavefront sensor (lower plots) for the following set of beam parameters: (\mbox{$\ax=\ay=\SI{5}{\milli\radian}$}, \mbox{$\px=\py=\SI{-1}{\milli\meter}$}, \mbox{$\cur=\SI{0.01}{\per\meter}$}, \mbox{$\ds=\SI{8.17}{\milli\meter}$}). The measured and offset-corrected values are \mbox{$(\axm=\SI{5.216}{\milli\radian}$}, \mbox{$\aym=\SI{4.928}{\milli\radian}$}, \mbox{$\pxm=\SI{-1.004}{\milli\meter}$}, \mbox{$\pym=\SI{-0.952}{\milli\meter}$}, \mbox{$\curm=\SI{0.020}{\per\meter}$}, \mbox{$\dsm=\SI{8.062}{\milli\meter}$}). Each point represents one of the rays traced in the simulation. Black circles mark weighted averages. Note the difference in axis scales between the measurement and sensor plane due to the magnification of the telescope. The apparent rotation of the distributions in the sensor plane indicate the presence of inter-axis crosstalk for centroid position and pointing angle.}
    			\label{fig:plot_intensity_phase_distributions}
    		\end{figure}

            The simulation of the beam diagnostic system is performed using Ansys Zemax OpticStudio in sequential ray tracing mode, excluding wave-optical effects, polarization, and spurious effects such as Fresnel reflections and scattering. The simulation starts in the measurement plane, where 621 rays with wavelength \SI{1064}{\nano\meter} are distributed on a rectangular grid with spacing \SI{0.5}{\milli\meter} (see \Cref{fig:plot_intensity_phase_distributions}). Initial convergence tests confirmed that the chosen sampling density provided sufficient accuracy, as further increasing the number of rays produced no significant changes in the results. The beam parameters $\Omega$ are configured by adjusting four values for each ray at position $(x,y)$ in the measurement plane: the relative intensity $I(x,y)=\exp(-8((x-\px)^2+(y-\py)^2)/\ds^2)\in[0,1]$, the phase $\varphi(x,y)$, and the initial pointing angles along the axes $x$ and $y$. To achieve the desired curvature $\cur$, the phase is implemented with an appropriately scaled quadratic profile~\cite{Hecht2018} centered at $(\px,\py)$. The resulting intensity and phase distributions in the measurement and sensor plane for a set of parameter values are illustrated in \Cref{fig:plot_intensity_phase_distributions}.
    
            The simulation traces the rays to the wavefront sensor plane, where the resulting intensity distributions (in position and angle space) and phase distributions are analyzed (see lower plots in \Cref{fig:plot_intensity_phase_distributions}). The measured parameters $\Omega_m$ are determined while accounting for both the telescope magnification $M$ and the axes inversion introduced by the imaging system. The magnification factor and sign reversals are applied to convert the measured values back to the beam measurement plane, ensuring correct parameter reconstruction in the coordinate system of the measurement plane. The pointing angles $\axm$ and $\aym$ are calculated using the weighted average of the angle distributions, following the formula $\axm=M\sum_i \axm(i) I(i) / \sum_i I(i)$, where $i$ indexes the rays and $\axm(i)$ represents the $x$-direction pointing angle of ray $i$ with relative intensity $I(i)$. Similarly, the centroid position parameters $\pxm$ and $\pym$ are derived from the weighted average $\pxm=M^{-1}\sum_i \pxm(i) I(i) / \sum_i I(i)$ of the position distributions $\pxm(i)$. The curvature $\cur$ is determined by fitting the measured phase distribution with the sum of Zernike polynomials up to order 5, following the ANSI standard order~\cite{Thibos00}. The beam diameter $\dsm$ is defined as $\dsm=(\ds_{x,m} + \ds_{y,m}) / 2$, which is the arithmetic mean of the beam diameters along the $x$ and $y$ axes. Along the $x$-axis, the beam diameter is calculated according to
            \begin{equation}
                \ds_{x,m} = 4M^{-1}\sqrt{\frac{\sum_i (\pxm(i)-\pxm)^2 I(i)}{\sum_i I(i)}}
            \end{equation}
            with an analogous expression for the $y$-axis.
    
	\section{Results}
        The simulation results are presented in two parts. First, a comprehensive analysis of crosstalk effects is performed, examining how variations in parameters affect the measurements of others and explaining the origins of these effects. Second, based on these findings, a mitigation method is developed to reduce measurement deviations caused by crosstalk.
        
    	\subsection{Crosstalk analysis}
        \label{subsection:crosstalk_analysis}            
            To investigate crosstalk effects in the beam diagnostic system, a comprehensive parameter sweep is conducted in the optical simulation. All six beam parameters are varied simultaneously within their respective measurement ranges (see \Cref{tab:measurement_ranges}). Five equidistant sampling points are selected across each parameter's measurement range, resulting in a total of \mbox{$5^6=15625$} unique parameter value combinations. The corresponding measured parameter values are recorded and analyzed for each combination. For the analysis, the figure-of-merit is the measurement deviation $\omega_m-\omega$ for $\omega\in\Omega$, i.e., the difference between the measured and set parameter value which determines the measurement accuracy. While systematic offsets between set and measured values are observed for all parameters, these are not considered critical for the system performance as they can be eliminated through a calibration procedure or post-processing of the measurement data.

\begin{figure}
                \centering
                \begin{subfigure}[b]{0.49\textwidth}
                    \centering
                    \includegraphics[width=\textwidth]{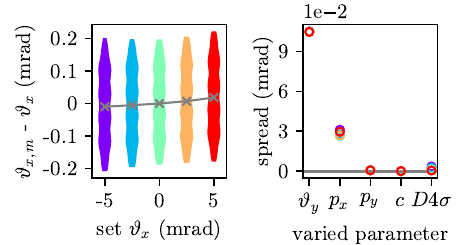}
                    \caption{pointing angle $\ax$}
                    \label{fig:plot_contributions_pointing_x_wfs_nice}
                \end{subfigure}
                \hfill
                \begin{subfigure}[b]{0.49\textwidth}
                    \centering
                    \includegraphics[width=\textwidth]{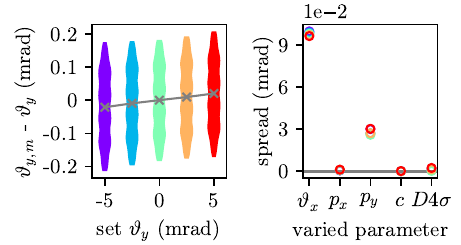}
                    \caption{pointing angle $\ay$}
                    \label{fig:plot_contributions_pointing_y_wfs_nice}
                \end{subfigure}

                \vspace{.5em}
                
                \begin{subfigure}[b]{0.49\textwidth}
                    \centering
                    \includegraphics[width=\textwidth]{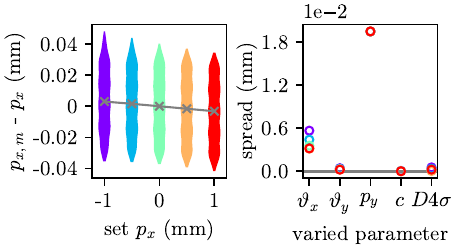}
                    \caption{centroid position $\px$}
                    \label{fig:plot_contributions_position_x_nice}
                \end{subfigure}
                \hfill
                \begin{subfigure}[b]{0.49\textwidth}
                    \centering
                    \includegraphics[width=\textwidth]{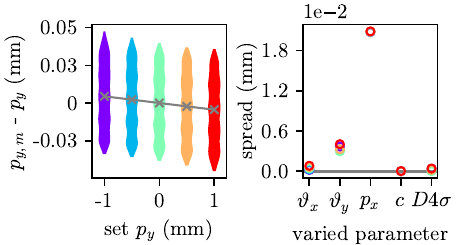}
                    \caption{centroid position $\py$}
                    \label{fig:plot_contributions_position_y_nice}
                \end{subfigure}
                
                \vspace{.5em}
                
                \begin{subfigure}[b]{0.49\textwidth}
                    \centering
                    \includegraphics[width=\textwidth]{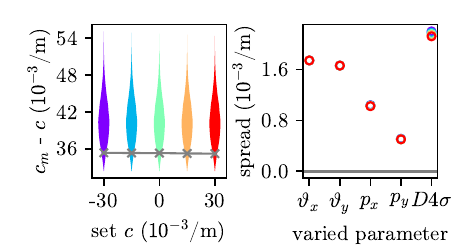}
                    \caption{curvature $\cur$}
                    \label{fig:plot_contributions_curvature_nice}
                \end{subfigure}
                \hfill
                \begin{subfigure}[b]{0.49\textwidth}
                    \centering
                    \includegraphics[width=\textwidth]{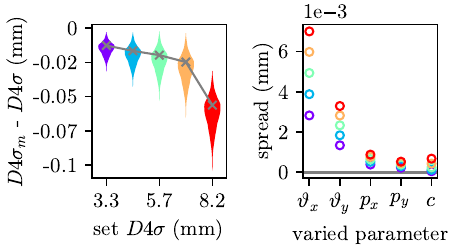}
                    \caption{diameter $\ds$}
                    \label{fig:plot_contributions_diameter_d95_nice}
                \end{subfigure}

                \vspace{.5em}
                
                \caption[Quantitative assessment of the strength and origins of parameter crosstalk]{Quantitative assessment of the strength and origins of parameter crosstalk. For each of the six parameters, the left panels present measurement deviations arising from crosstalk across all sampled parameter combinations. The distribution of the measurement deviation is shown as a violin plot for each set parameter value. The measured values for pointing angles and centroid positions are shown after subtracting an offset value (obtained at nominal parameters), simulating a typical calibration procedure. Gray crosses mark the points where all other set parameters are nominal. The right panels enable identification of individual parameter contributions to the observed crosstalk by displaying the spread of the measurement deviation when varying single parameters while maintaining others at nominal values. The color scheme indicates the set value of each investigated parameter and is consistent between corresponding left and right panels.}
                \label{fig:contributions}
            \end{figure}

            The evaluation results, providing a quantitative assessment of the strength and origins of parameter crosstalk, are summarized in \Cref{fig:contributions}. For each measured parameter, the distribution of measurement deviations across all sampled parameter combinations is presented as a violin plot, grouped by set parameter values. A complementary analysis quantifies individual parameter contributions to the observed crosstalk by examining the spread of measurement deviations. This spread is defined as the standard deviation of all measurement deviations observed when varying a single parameter while maintaining all others at their nominal values. A spread close to zero indicates that the isolated variation of the corresponding parameter does not substantially influence the measurement outcome, while non-zero spread values reveal crosstalk between the set parameter and the measured parameter.

            The crosstalk analysis reveals distinct patterns of parameter interdependencies caused by the optical elements of the beam diagnostic system. The observed effects can be systematically analyzed for each measured parameter, providing insights into the underlying mechanisms and their contributions to measurement deviations.
            
            The pointing angle measurements (\Cref{fig:plot_contributions_pointing_x_wfs_nice,fig:plot_contributions_pointing_y_wfs_nice}) exhibit three primary crosstalk mechanisms, with the inter-axis coupling emerging as the dominant contribution by a significant margin. This coupling between $x$ and $y$ pointing angles manifests as a mutual influence where changes in one pointing angle affect the measurement of the other. The observed effect primarily stems from the beam pickup wedge, which couples the two angular coordinates. This is validated through analytical calculations of the reflection at the beam pickup wedge (see \Cref{sec:app:analyticalcrosstalk}), yielding predicted standard deviations of \SI{10e-2}{\milli\radian} for each axis. This is close to the observed values of \SI{10.5e-2}{\milli\radian} and \SI{9.8e-2}{\milli\radian} for the $x$- and $y$-axis, respectively. The rotating wedges have negligible impact on this coupling due to near-perpendicular angles of incidence on the wedged surfaces. Two additional, albeit significantly weaker crosstalk mechanisms are present: First, pointing measurements show sensitivity to beam position variations, stemming from imperfect telescope alignment at the operating wavelength of \SI{1064}{\nano\meter}. Second, a correlation with beam diameter is observed through increasing aberration effects for off-axis rays as the beam diameter grows.
            
            Centroid position measurements (\Cref{fig:plot_contributions_position_x_nice,fig:plot_contributions_position_y_nice}) demonstrate similar characteristic dependencies, with the inter-axis coupling again being the primary effect. This coupling between $x$ and $y$ centroid positions originates from the beam pickup wedge, which affects not only the pointing angles but also introduces positional coupling. The effect is clearly visible in the position-space plots at the sensor plane in \Cref{fig:plot_intensity_phase_distributions}, where the positional coupling manifests itself as a slight rotation of the grid of rays. Like the pointing measurements, the centroid position measurements are similarly affected by imperfect telescope alignment and off-axis aberrations.

            Beam diameter measurements (\Cref{fig:plot_contributions_diameter_d95_nice}) reveal crosstalk with multiple parameters. The most significant effect is the sensitivity to pointing variations, with a smaller contribution from position variations. These effects can be traced to lateral offsets in the $x$-direction at the first lens, originating from the rotating wedge configuration. This offset is notably larger in the $x$-direction than the $y$-direction, leading to asymmetric behavior. The effect manifests through off-axis aberrations that become increasingly pronounced with larger beam diameters. Furthermore, the diameter measurements show sensitivity to beam curvature variations, attributed to imperfect alignment between the sensor and image planes, causing apparent diameter changes when the beam curvature is varied.
            
            Curvature measurements (\Cref{fig:plot_contributions_curvature_nice}) exhibit the highest susceptibility to crosstalk, responding to all aforementioned effects, making it the most challenging parameter to measure accurately. The distribution width of curvature measurement deviations reaches \SI{20e-3}{\per\meter}, comparable to the full measurement range of $\pm\SI{30e-3}{\per\meter}$. Without appropriate crosstalk mitigation, this limits the measurement accuracy to approximately one third of the measurement range. Additionally, the curvature measurements show a systematic offset of about \SI{40e-3}{\per\meter}, which is caused by the imperfect telescope alignment at the operating wavelength.

            The magnitude of these measurement deviations, particularly evident in the curvature measurement, demonstrates that accurate beam characterization requires systematic crosstalk mitigation strategies.

    	\subsection{Crosstalk mitigation}
            Following the analysis of the various crosstalk mechanisms present in the system, this section presents a mathematical framework to correct for these effects to some extent and recover the beam parameters from the measured values with improved accuracy. The approach consists of developing mathematical expressions that approximately describe the dependencies of measured parameters on set parameters, then solving these equations for the set parameters to reconstruct them from a set of measured values.
        
            Due to the presence of crosstalk, each measured parameter must, in general, be expressed as a function of all set parameters. For example, the measured pointing angle in the $x$-direction can be described by an unknown function $g_{\ax}$ that maps the set parameters to the measured value:
            \begin{equation}
                \axm = g_{\ax}(\Omega)
            \end{equation}
            In the absence of crosstalk and systematic offsets, this function would simplify to $g_{\ax}(\Omega) = \ax$. While the functional relationships between set and measured parameters are unknown, Taylor series expansions provide a powerful approach to approximating them. Without requiring knowledge of the explicit form, a Taylor series can represent a smooth function in the neighborhood of a point as an infinite sum of terms calculated from the function's derivatives, where each successive term involves higher derivatives and higher powers~\cite{lang1996calculus}. In the neighborhood of the nominal parameter values, each of the functions $g_\omega(\Omega)$ with $\omega\in\Omega$ can be expressed as
            \begin{equation}
                g_\omega(\Omega) = \sum_{\omega'\in\Omega}\sum_{|\alpha| \geq 0} \frac{D^\alpha g_{\omega'}\rvert_{\Omega_\mathrm{nom}}}{\alpha!}\Omega^\alpha,
            \end{equation}
            with the multi-index $\alpha=(\alpha_1,...,\alpha_6)$, $D^\alpha g_\omega = \partial^{|\alpha|}g_\omega / \partial x_1^{\alpha_1}...\partial x_6^{\alpha_6}$, $|\alpha|=\alpha_1+...+\alpha_6$, $\alpha!=\alpha_1!...\alpha_6!$, $\Omega^\alpha=\omega_1^{\alpha_1}...\omega_6^{\alpha_6}$, and $\Omega_\mathrm{nom}$ representing the nominal parameter values. This general form presents two significant practical challenges. First, the multi-dimensional nature of the expansion demands extensive calibration data with densely sampled, simultaneous variations of all parameters to approximate the derivatives. Second, the infinite number of terms in the series expansion requires determining an infinite number of coefficients from the calibration data. Two key approximations address these limitations.
            
            The first approximation neglects cross terms that depend on more than one parameter, such as terms of the form $\omega_1\omega_2$ for any \mbox{$\omega_1,\omega_2\in\Omega$}. More generally, the approximation excludes terms of the form $\omega_1^{\alpha_1}...\omega_6^{\alpha_6}$ with non-zero exponents for at least two parameters. These cross terms represent coupled effects where the influence of one parameter on the measurement depends on the value of another parameter. For example, a cross term between pointing angle and beam diameter would indicate that the impact of pointing angle variations on the measurement changes with the beam size. This approximation effectively reduces the series expansion to a sum of one-dimensional Taylor series
            \begin{equation}
                g_\omega(\Omega) \approx \sum_{\omega'\in\Omega} \sum_{n=0}^\infty \frac{1}{n!}\frac{\partial^n g_\omega}{\partial \omega'^n} \biggr\rvert_{\Omega_\mathrm{nom}} \omega'^n.
            \end{equation}
            This simplification substantially reduces the number of required coefficients and simplifies the calibration procedure, as single parameter sweeps are sufficient for the determination of the coefficients.

            \begin{figure}[t]
    			\centering
    			\includegraphics[width=.5\textwidth]{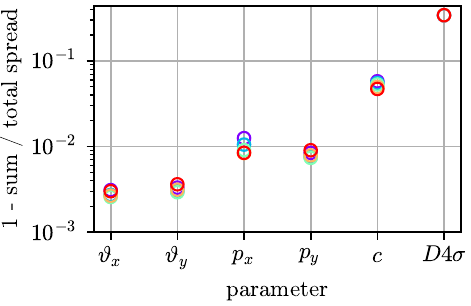}
    			\caption[Comparison of the sum of spreads to the total spread for each parameter]{Comparison of the sum of spreads to the total spread for each parameter. The sum of spreads is the root-mean-square sum of individual spread values when varying each parameter separately, while the total spread is obtained when varying all parameters simultaneously. Each color represents one of the sampled set values for each parameter and the color mapping is the same as in \Cref{fig:contributions}.}
    			\label{fig:plot_spread_sum}
    		\end{figure}
            
            The validity of this approximation can be assessed by comparing two different analyses. First, measuring how each parameter's variation affects the measurement individually while keeping other parameters fixed at their nominal values. Second, measuring the total variation when all parameters change simultaneously. By comparing the root-mean-square sum of the individual variations to the total variation observed across all parameter combinations, the significance of cross terms can be evaluated. If cross terms are negligible, the total variation should approximately equal the sum of individual variations, following the principle that variances of independent effects add linearly~\cite{hogg2005introduction}. As shown in \Cref{fig:plot_spread_sum}, this is indeed the case for most parameters: for pointing angle and centroid position, the agreement is above \SI{98}{\percent}, the curvature exhibits less than \SI{10}{\percent} discrepancy, while beam diameter measurements show a potential limitation of the approximation with about \SI{40}{\percent} deviation.
            
            While this analysis supports neglecting cross terms for the chosen measurement ranges, their contribution is expected to become more significant when sampling a larger parameter space. As parameters deviate further from their nominal values, coupling effects between parameters likely become more relevant, potentially limiting the validity of this approximation. The significance of such cross terms can be assessed more quantitatively by computing numerical derivatives from data that captures simultaneous variations in multiple parameters.

            \begin{figure}[t]
                \centering
                \begin{subfigure}[b]{0.32\textwidth}
                    \centering
                    \includegraphics[width=\textwidth]{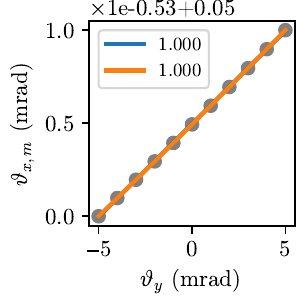}
                    \caption{$\axm$ vs. $\ay$}
                    \label{fig:plot_fit_pointing_x_pointing_y}
                \end{subfigure}
                \hfill
                \begin{subfigure}[b]{0.32\textwidth}
                    \centering
                    \includegraphics[width=\textwidth]{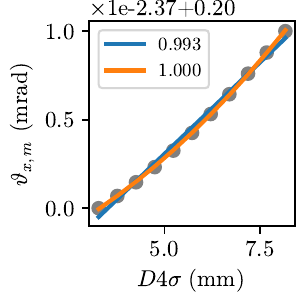}
                    \caption{$\axm$ vs. $\ds$}
                    \label{fig:plot_fit_pointing_x_diameter_d95}
                \end{subfigure}
                \hfill
                \begin{subfigure}[b]{0.32\textwidth}
                    \centering
                    \includegraphics[width=\textwidth]{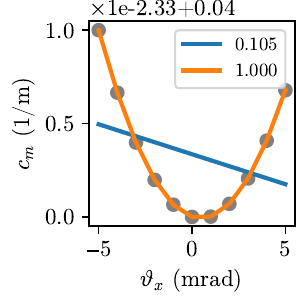}
                    \caption{$\curm$ vs. $\ax$}
                    \label{fig:plot_fit_curvature_pointing_x}
                \end{subfigure}
                
                \vspace{.5em}
                
                \caption[Exemplary results of single parameter sweeps and fits with first and second-order polynomials]{Exemplary results of single parameter sweeps and fits with first- and second-order polynomials (blue and orange lines, respectively). For each parameter, 11 equidistant points are sampled. The R squared values in the legends indicate the goodness of the fits to the data~\cite{Lewis-Beck1990-ol}. The full set of all 36 plots is provided in \Cref{sec:app:fulldatacorrection}.}
                \label{fig:plot_fits}
            \end{figure}
            
            Following the elimination of cross terms, the second approximation aims to reduce the number of orders in the series expansion. To determine both the appropriate truncation order and the coefficients of the Taylor expansion, systematic single-parameter sweeps are performed. Each sweep comprises eleven equidistant sampling points spanning the full measurement range, with all other parameters held constant at their nominal values. The relationships between set and measured parameters, exemplarily shown in \Cref{fig:plot_fits}, reveal varying degrees of nonlinearity. For instance, $\axm$ exhibits a clear linear dependence on $\ax$, while the response of $\curm$ to variation of $\ax$ is quadratic. Second-order polynomial fits describe the data well across all parameter combinations (see \Cref{sec:app:fulldatacorrection}). Consequently, the Taylor series is truncated after the quadratic term, yielding
            \begin{equation}
                \label{eq:correction_equations}
                g_\omega(\Omega) \approx a_{\omega} + \sum_{\omega'\in\Omega} \left(b_{\omega,\omega'} \omega' + c_{\omega,\omega'} \omega'^2 \right),
            \end{equation}
            with $a_{\omega}, b_{\omega,\omega'}, c_{\omega,\omega'}$ the polynomial coefficients. The constant term $a_{\omega}$ describes a systematic offset, which is the same for all varied parameters $\omega'$, and is therefore extracted from the sum. The values of the \mbox{$6^2 \times 2 + 6 = 78$} coefficients are determined from the second-order polynomial fits shown in \Cref{fig:plot_fits_all} in \Cref{sec:app:fulldatacorrection}.

            \begin{figure}[t]
                \centering
                \begin{subfigure}[b]{0.32\textwidth}
                    \centering
                    \includegraphics[width=\textwidth]{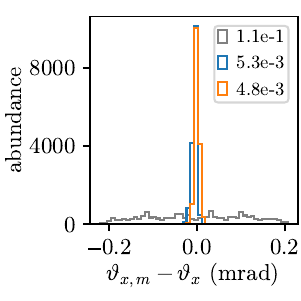}
                    \caption{pointing angle $\ax$}
                    \label{fig:plot_correction_pointing_x}
                \end{subfigure}
                \hfill
                \begin{subfigure}[b]{0.32\textwidth}
                    \centering
                    \includegraphics[width=\textwidth]{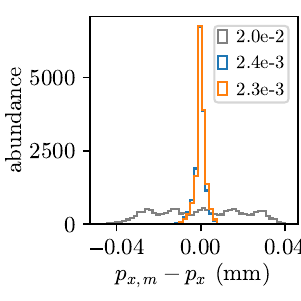}
                    \caption{centroid position $\px$}
                    \label{fig:plot_correction_position_x}
                \end{subfigure}
                \hfill
                \begin{subfigure}[b]{0.32\textwidth}
                    \centering
                    \includegraphics[width=\textwidth]{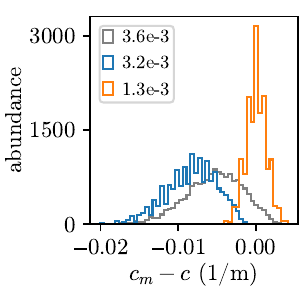}
                    \caption{curvature $\cur$}
                    \label{fig:plot_correction_curvature}
                \end{subfigure}

                \vspace{.5em}
                
                \begin{subfigure}[b]{0.32\textwidth}
                    \centering
                    \includegraphics[width=\textwidth]{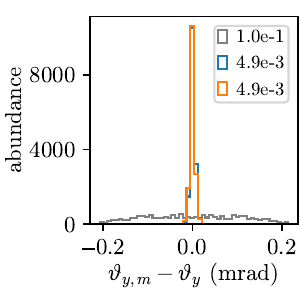}
                    \caption{pointing angle $\ay$}
                    \label{fig:plot_correction_pointing_y}
                \end{subfigure}
                \hfill
                \begin{subfigure}[b]{0.32\textwidth}
                    \centering
                    \includegraphics[width=\textwidth]{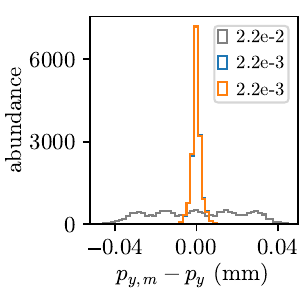}
                    \caption{centroid position $\py$}
                    \label{fig:plot_correction_position_y}
                \end{subfigure}
                \hfill
                \begin{subfigure}[b]{0.32\textwidth}
                    \centering
                    \includegraphics[width=\textwidth]{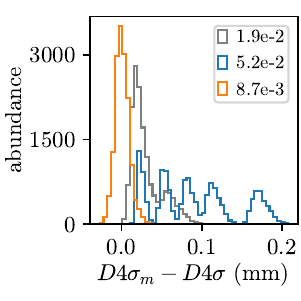}
                    \caption{diameter $\ds$}
                    \label{fig:plot_correction_diameter_d95}
                \end{subfigure}

                \vspace{.5em}
                
                \caption[Mitigation of crosstalk]{Mitigation of crosstalk. For each parameter, the histogram shows the distribution of the measurement deviation of the uncorrected measurement data (gray) and with crosstalk mitigation based on first- or second-order polynomials applied (blue or orange, respectively). To enable comparability, the offset in the uncorrected measurement data (see \Cref{fig:contributions}) is eliminated by subtracting the measured value at nominal parameter settings. Legends indicate the standard deviations of the distributions in units as indicated on the axes labels. For the pointing angle and centroid position parameters, the mitigation results based on first- or second-order polynomials yield mostly congruent distributions.}
                \label{fig:corrections}
            \end{figure}
            
            The crosstalk mitigation method is tested by numerically solving the set of six coupled equations $\{\omega_m = g_\omega(\Omega)|\omega\in\Omega\}$ for the set parameters $\Omega$. For each of the 15625 parameter combination from the parameter sweep discussed in \Cref{subsection:crosstalk_analysis}, the measured values are input into these equations to obtain corrected values, partially accounting for the crosstalk effects and providing improved measurement accuracy. The results of this evaluation are summarized in \Cref{fig:corrections}, where the distribution of the measurement deviation for each parameter with and without the correction is shown. For comparison, the results of a correction based on first-order polynomials ($c_{\omega,\omega'}=0$ in \Cref{eq:correction_equations}) are also included. Further investigations using polynomials with order higher than two yielded no significant improvement.

            The application of this method yields two key benefits. First, the systematic offset present in the uncorrected data is eliminated through the inclusion of constant terms in \Cref{eq:correction_equations}, leading to enhanced measurement trueness~\cite{iso5725-1}. Second, the spread of measurement deviations is substantially reduced, indicating improved precision~\cite{iso5725-1}. This improvement is particularly pronounced for the pointing angle and centroid position parameters, where the spread decreases by more than one order of magnitude. The spread of the curvature and diameter are also decreased, though to a lesser extent with precision improvements by factors between 2 and 3. This difference likely stems from the current mitigation method's neglect of cross terms, which appear to be more significant for the latter two parameters, as evidenced by the analysis shown in \Cref{fig:plot_spread_sum}.
			
	\section{Conclusion and Outlook}
        The results of this study demonstrate both the potentially significant impact of crosstalk effects in multi-parameter laser beam diagnostic systems and the effectiveness of a systematic mitigation approach in reducing these effects. The comprehensive analysis revealed that without mitigation, crosstalk severely limits the attainable measurement accuracy in the investigated setup. In particular, the curvature measurements showed deviations comparable to the measurement range itself, highlighting the critical need for crosstalk mitigation in the practical application discussed herein. The demonstrated mitigation method, based on Taylor series expansions, achieved substantial improvements in measurement accuracy.
        
        However, several aspects warrant further investigation to advance this work toward practical implementation. First, the current mitigation method could be enhanced by including cross terms in the Taylor expansion. While this would significantly increase the calibration effort due to the need for simultaneous variation of multiple parameters, it might provide better mitigation for parameters like curvature and diameter measurements, where the analysis suggested more significant contributions from coupled effects. For 6 parameters, the number of cross term parameter combinations that can be investigated is given by the binomial coefficient~\cite{cameron1994combinatorics} $\binom{6}{2}=15$.
        
        A crucial limitation of the present study is its focus on the nominal optical system without considering tolerances of optical elements and their mounting. Future research should investigate how these tolerances affect the mitigation method's performance and how the mitigation quality derived from the nominal system deteriorates when applied to systems with tolerances. This could be accomplished through Monte-Carlo simulations incorporating tolerance distributions of optical and mechanical components.
        
        In addition, future studies should include modeling of the wavefront sensor. The current simulation neglects effects such as finite pixel size, electronic noise, dynamic range limitations, and the discretization of the wavefront by the lenslet array or phase-reconstruction algorithm. Including these factors would provide more realistic predictions of the achievable measurement accuracy in practical implementations and might reveal additional limitations or requirements for the mitigation method.
        
        Furthermore, while the ray-optical simulations presented here provide valuable insights into the system's behavior, they exclude wave-optical effects such as diffraction and interference. These effects are relevant, for example, close to the focal point of the lenslets of a Shack-Hartmann sensor and the sensor plane of a multi-wave lateral shearing interferometer.
        
        The ultimate validation of the mitigation method requires experimental verification. Laboratory measurements with an actual implementation of the diagnostic system would reveal practical challenges not captured by the simulations and provide crucial feedback for refining the mitigation approach. Such measurements could also help establish uncertainty budgets for the corrected measurements, taking into account all relevant error sources in a real system.

        The approach developed here draws interesting parallels to established crosstalk mitigation methods in array detector spectrometers, in which stray light caused by effects such as surface scattering and Fresnel reflections can create crosstalk between wavelength channels~\cite{Metrologiastray}. A common method for spectrometers characterizes the instrument response matrix using known monochromatic inputs and recovers true signals through matrix inversion~\cite{Zong:06,10.1364/AO.53.004313,barlier2014stray}. While the physical origins differ from the system discussed here - stray light in spectrometers versus parameter coupling in optical imaging systems - the underlying principle of systematic characterization and mathematical modeling is similar. The Taylor series approach presented here extends beyond simple matrix inversion, which is limited to linear relationships, by including non-linear terms. Furthermore, Taylor series allow for the inclusion of cross terms of arbitrary order, enabling the description of parameter coupling effects and providing a flexible framework for crosstalk mitigation in complex optical systems.
        
    \newpage
    \section{Appendix}
    \begin{appendices}
        \section{Details of the optical setup}
        \label{sec:app:opticalsystem}

            \Cref{tab:optical_system} presents the optical surface parameters used in the sequential Zemax simulation of the beam diagnostic system shown in \Cref{fig:system_layout}.
            
            \begin{table}[htbp]
                \centering
                \small
                \begin{tabular}{|S[table-format=2.0]|l|l|S[table-format=3.2]|S[table-format=2.4]|l|S[table-format=2.1]|}
                    \hline
                    {\makecell{Surface\\Number}} & Comment & \makecell{Zemax Surface Type} & {\makecell{Radius of\\Curvature\\(\si{\milli\meter})}} & {\makecell{Thickness\\(\si{\milli\meter})}} & Material & {\makecell{Radius\\(\si{\milli\meter})}} \\
                    \hline
                    0 & OBJECT & Standard & & & & \\
                    1 & measurement plane & Standard & & & & 5 \\
                    2 & wavefront setting & Zernike Fringe Phase &  & & & \\
                    3 & spacing & Standard &  & 25 & & \\
                    4 & \SI{45}{\degree} rotation & Coordinate Break & & & & \\
                    5 & beam pickup & Tilted & & & MIRROR & 12.7 \\
                    6 & \SI{45}{\degree} rotation & Coordinate Break & & & & \\
                    7 & spacing & Standard &  & -15 & & \\
                    8 & rotation $y$: \SI{-102.8}{\degree} & Coordinate Break & & & & \\
                    9 & rotating wedge 1 & Standard &  & -5 & N-BAF10 & 12.7 \\
                    10 & rotating wedge 1 & Tilted & & & & 12.7 \\
                    11 & rotation $y$: \SI{46.0}{\degree} & Coordinate Break & & & & \\
                    12 & spacing & Standard &  & -3 & & \\
                    13 & rotating wedge 2 & Tilted & & -5 & N-SF6HT & 12.7 \\
                    14 & rotating wedge 2 & Standard &  & & & 12.7 \\
                    15 & rotation $y$ reset & Coordinate Break & & & & \\
                    16 & spacing & Standard & & -10 & & \\
                    17 & AC254-100-C & Standard & -32.14 & -6.5 & N-BAF10 & 12.7\\
                    18 & AC254-100-C & Standard & 38.02 & -1.8 & N-SF6HT & 12.7\\
                    19 & AC254-100-C & Standard & -93.54 & -90.2244 & & 12.7\\
                    20 & field stop & Standard &  & & &\\
                    21 & spacing & Standard &  & -41.3599 & & \\
                    22 & AC254-050-C & Standard & 194.54 & -1.8 & N-SF6HT & 12.7 \\
                    23 & AC254-050-C & Standard & -25.88 & -9 & N-LAK22 & 12.7\\
                    24 & AC254-050-C & Standard & 22.94 & -64.3085 & & 12.7 \\
                    25 & IMAGE & Standard &  & & & \\
                    \hline
                \end{tabular}
                \caption{Surface parameters of the optical setup.}
                \label{tab:optical_system}
            \end{table}

        \newpage
        \section{Analytical calculation of inter-axis pointing crosstalk}
        \label{sec:app:analyticalcrosstalk}
            In the coordinate system defined in \Cref{fig:system_layout}, the wave vector $\bm{k}$ of a ray with pointing angles $\ax$ and $\ay$ incident on the wedged beam splitter is given by
            \begin{equation}
                \bm{k} = \begin{pmatrix} k_1 \\ k_2 \\ k_3 \end{pmatrix} = \begin{pmatrix} -\sin(\ax) \\ -\sin(\ay) \\ \sqrt{\cos^2(\ax)-\sin^2(\ay)} \end{pmatrix},
            \end{equation}
            and the surface normal vector $\bm{n}$ of the wedged surface with wedge angle $\beta=\SI{20}{\milli\radian}$ is
            \begin{equation}
                \bm{n} = \begin{pmatrix} n_1 \\ n_2 \\ n_3 \end{pmatrix} = \begin{pmatrix} -\sin(\beta) \\ \cos(\beta)/\sqrt{2} \\ -\cos(\beta)/\sqrt{2} \end{pmatrix}.
            \end{equation}
            Using the law of reflection in vectorial form \cite{Bhattacharjee_2005}, the wave vector of the reflected ray $\bm{k_r}$ is calculated as
            \begin{equation}
                \bm{k_r} = \bm{k} - 2\frac{\bm{k}\cdot\bm{n}}{\left\lVert\bm{n}\right\rVert^2}\bm{n},
            \end{equation}
            where $\cdot$ denotes the dot product and $\left\lVert\right\rVert$ the Euclidean norm. The pointing angles of the reflected ray are obtained from the arc cosine of the components of $\bm{k_r}$. The standard deviation of the measured $\ay$ angles for 5 different $\ax$ inputs yields \SI{1.000e-1}{\milli\radian}, and vice versa for $\ay$ inputs yielding \SI{9.997e-2}{\milli\radian} in $\ax$ measurements.

            A similar analysis is conducted for one of the rotating wedges with the wedged surface aligned to the $x$-axis and perpendicular incidence. In this case, the surface normal vector of the wedged surface is described by
            \begin{equation}
                \bm{n} = \begin{pmatrix} -\sin(\beta) \\ 0 \\ -\cos(\beta) \end{pmatrix},
            \end{equation}
            with $\beta=\SI{50}{\milli\radian}$, and the wave vector of the transmitted ray $\bm{k_t}$ is calculated using the vectorial form of the law of refraction~\cite{Miks:12}
            \begin{equation}
                \bm{k_t} = \sqrt{1-\mu^2(1-(\bm{n}\cdot\bm{k})^2)}\bm{n} + \mu(\bm{k}-(\bm{n}\cdot\bm{k})\bm{n})
            \end{equation}
            with $\mu=n_1/n_2$ the ratio of the refractive indices of the medium of the incident ray $n_1$ and transmitted ray $n_2$. The calculation accounts for two refractions: one at the front surface and another at the wedged back surface or the rotating wedge. The standard deviation of transmitted ray angles for 5 different $\ax$ inputs yields \SI{0}{\milli\radian}, and for $\ay$ inputs yields \SI{1.038E-04}{\milli\radian}.
        
        \newpage
        \section{Full evaluation of single parameter sweeps}
        \label{sec:app:fulldatacorrection}
        
    		\begin{figure}[htbp]
    			\centering
    			\includegraphics[width=1\textwidth]{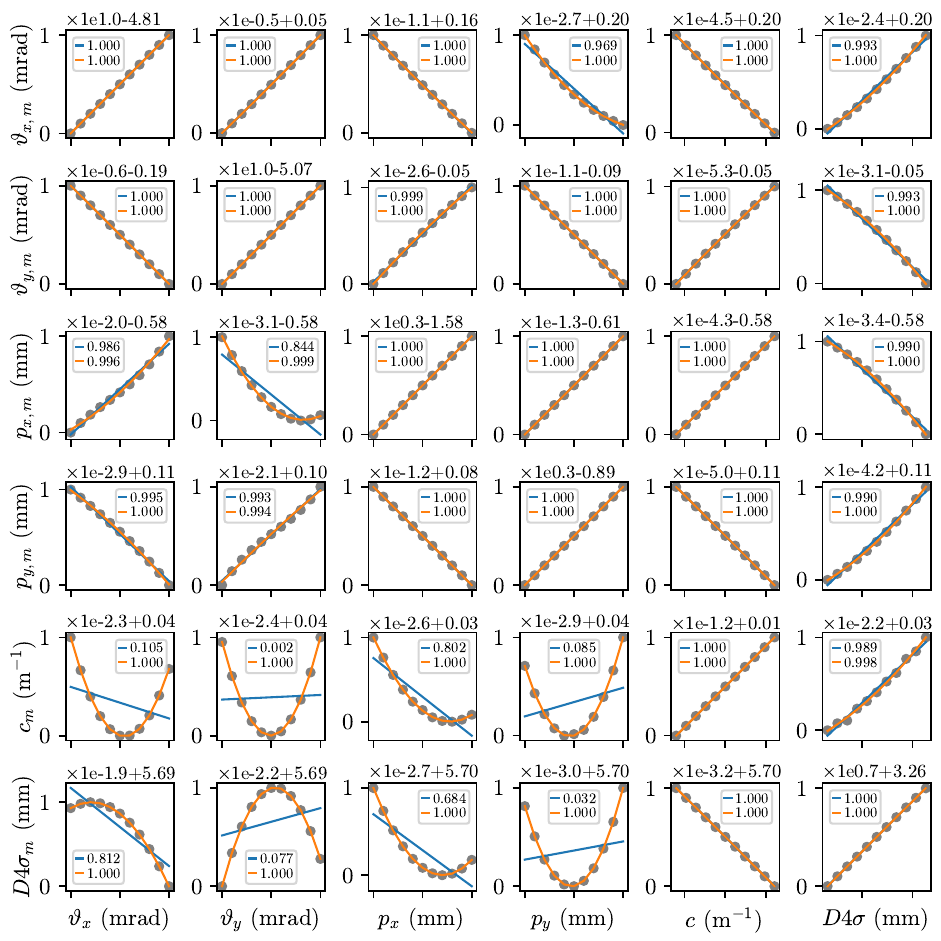}
    			\caption[Full results of single parameter sweeps and fits with first- and second-order polynomials]{Full results of single parameter sweeps and fits with first- and second-order polynomials. The plots are arranged in a $6 \times 6$ grid, where each column shows sweeps of a different beam parameter: beam pointing angles, centroid positions, curvature, and beam diameter. Each row represents a different measured parameter. The specific scaling of each plot is shown in the plot titles. Blue and orange lines represent first and second-order polynomial fits respectively, with their R squared values shown in the legends. The axes labels for all plots in a row are given on the leftmost plot, and the parameter being swept is labeled on the bottom row of plots.}
    			\label{fig:plot_fits_all}
    		\end{figure}
    
    \end{appendices}

    \newpage
    \section*{}\vspace{-\baselineskip}
    \subsection*{Disclosures}
        The author is an employee of Hensoldt Optronics GmbH, where the work described herein has been performed. The author declares no personal financial interests, patent applications, or other conflicts of interest that could influence the presented research.
    
    \subsection*{Code, Data, and Materials Availability}     
        The simulation data, Zemax optical design file, and Python analysis code underlying the results presented in this paper are not publicly available at this time but may be obtained from the author upon request.
    
    \subsection*{Acknowledgments}
        The author thanks Jürgen Werner for insightful discussions and for carefully reading the manuscript. The manuscript was checked for grammar and spelling errors using the Claude large language model (Anthropic, version 3.5 Sonnet, October 2024).
    
    \bibliography{literature}
    \bibliographystyle{spiejour}
    
    \vspace{2ex}\noindent\textbf{Dr. Benjamin Nagler} works as a system engineer at Hensoldt Optronics GmbH in Oberkochen/Germany. He received his diploma and PhD degrees in physics from the Technical University of Kaiserslautern in 2016 and 2020, respectively. His current work focuses on the development of laser beam metrology systems for the semiconductor manufacturing industry.
        
%
    
    \end{spacing}
    
\end{document}